# Knowledge, Trust, Security and Covertness In Massively Distributed Social Platforms: An Epistemic Networks Approach

Mihnea C. Moldoveanu and Joel A.C. Baum

University of Toronto

ABSTRACT

Social networks – which are often seeded, crystallized, grown and structured by large-scale social platforms like Facebook, LinkedIn, Slack, Twitter, WhatsApp, Instagram and WeChat – exhibit complicated *epistemic structures and dynamics* that arise from the nature of *interactive knowledge* that users/participants possess, share and shape: what users know about what other users know, about what other users know they know, and so forth, plays an important role in channeling user behavior on the platform as well as the structuring and dynamics of the social network the platform engenders. We use a novel approach to characterize knowledge states and flows in social networks – based on epistemic networks, or *epinets* – to study the relationship between the structure of platforms and the structure of social networks that forms on them.

## Epinets as Weakly Constrained and Expressively Rich Representations of Social Networks

We introduced epistemic networks – epinets [Moldoveanu and Baum, 2014; 2021] – as a tool for modeling the states of knowledge and interactive knowledge in social networks. An epinet connects people and propositional beliefs in a way that allows us to quickly represent *who knows what*, *who knows whom*, and *who knows what is known by whom* in ways that do not rely on the standard state spaces of epistemic game theory and that impose minimal structure on the propositions that people know, or think they know.

Here, we apply epistemic networks to the study of the social networks that are seeded, crystallized and shaped by *social platforms* that endow users with specific decision rights over elementary actions (e.g., 'friend/unfriend', 'share video', 're-tweet', etc.) and specific ways in which information about users is displayed and shared with other users (e.g., 'who sees I know whom?', 'who sees what I have shared with whom?' etc.). The epinets modeling tool allows for both the analysis and understanding of existing social structures spawned by massively distributed social platforms and for the design of new information sharing platforms that are likely to create certain network dynamics.

## Epistemic Network Building Blocks

To characterize the epinets that popular social platforms (e.g., Facebook, Twitter, Slack, LinkedIn) are most likely to induce, we need to introduce several epistemic modeling tools that allow us to represent what actors know and how they know it.

### Epistemic network actor states

The core building blocks of an epinet are not only individual agents and the propositions relevant to their interaction(s) but also their epistemic states *vis-à-vis* these propositions and each other's beliefs [see Figure 1]:

*Knowledge (K)*: $AkP$ '*A* knows *P*', for some agent *A* and some proposition *P*. If agent *A* knows, for instance, *P*='*a judgment in favor of the defendant was entered today*' then agent *A* will, *ceteris paribus*, act as if *P* is true in those cases in which (a) *P* is relevant and (b) *A* sees *P* as relevant (in which case we will say that *P* is salient to *A)*. $k$ is a simple binary relation (either $AkP$ or $\sim AkP$) between an agent and a proposition *P* that is characterized by the following necessary conditions: (a) *A* believes *P*, and (b) *P* is true[1]. *Belief* is a weaker state: $AbP$ ('*A* believes *P*') simply represents the weaker (first prong) of the conjunction above.

*Awareness and Awareness of Level n* ($k^n$): '*A* knows *P* & knows she knows *P*' is a state we will refer to as awareness. '*A* knows she knows *P*, and so forth, to *n* levels': $AkP, Ak(AkP), Ak(Ak(AkP)), Ak(....AkP)...)$, abbreviated $Ak^nP$ generalizes awareness to a state in which *A* knows *P*, knows that she knows it, and so forth. This state relates to an agent's introspective insight into the contents of her knowledge. It is possible that *A* knows *P* but does not know she knows *P* (i.e., *AkP&~(Ak(AkP))*). Call this state *unawareness* of *P*.

---

[1] These are not *sufficient* conditions for knowledge [Gettier, 1963]. *A* may believe *P*='*There is a quarter in my pocket*' on the basis of the vague sensation of a weight in his left pocket, and there is, indeed, a quarter in *A*'s pocket, but it is in his right pocket, not his left one. In this case, *A* does *not* know *P*, even though he has a valid reason for a true belief. What counts is having the *right* reason for holding the belief, i.e. a reason that may be causally related to the proposition expressed by the belief being true.

*Ignorance*: '*A* does not know *P* but knows she does not know *P*': *~(AkP)&Ak(~AkP)*. Ignorance of *P* (which can also understood as uncertainty about *P*) is such that *A* pays attention to the values of variables relevant to the truth value of *P*, but does not know those values or the truth value of *P*.

*Oblivion*: '*A* does not know *P*, does not know she does not know it, and so forth': *~(AkP)&~(Ak(~AkP))&* …. In this state, *A* does not know *P* nor is *heedful*, in any way, of *P*, or of information that could inform *A* about the truth value of *P*. She does not pay attention to variables or experiences that could impact the truth value of *P*. Whereas an ignorant *A* will raise questions about *P* when *P* is relevant, and an *A* who is level-2-aware of her ignorance of *P* will raise questions about *P* when *P* is relevant, an *A* who is oblivious of *P* will do neither.

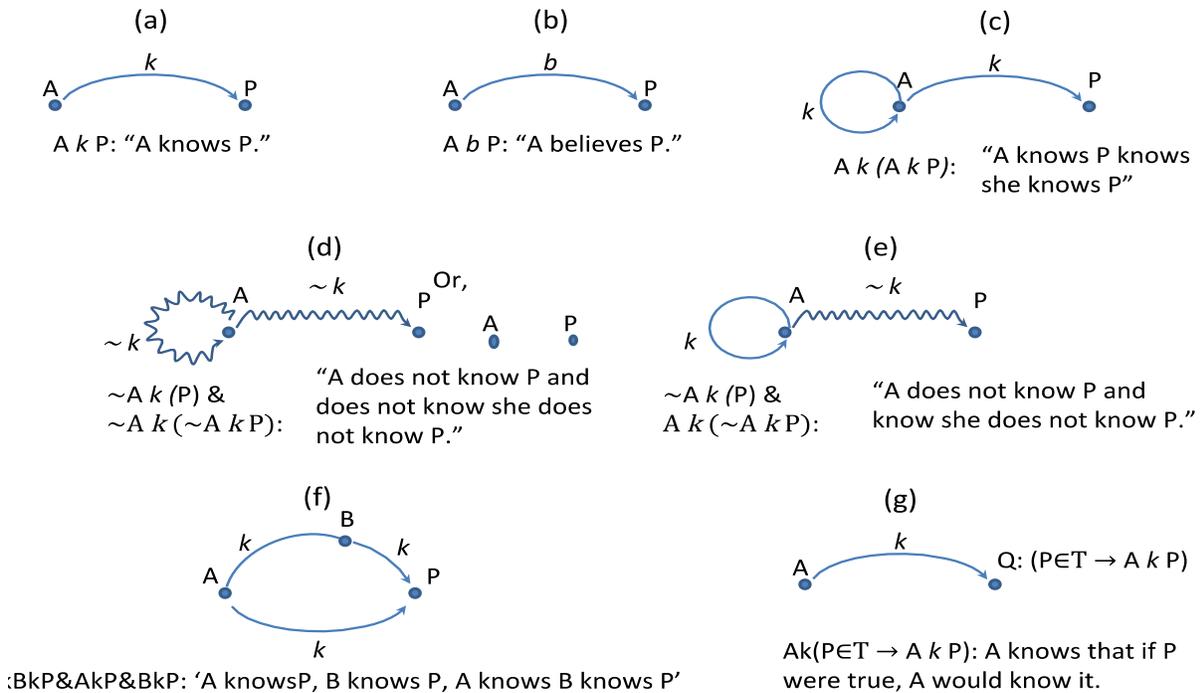

Figure 1. Epistemic State Representations for Dyadic Epinets.

Our definition of knowledge states can be augmented by an *epistemic qualifier* that refers to an agent's epistemic state. One may believe a proposition *P* is true to a certain degree *w*, but, within the set of beliefs that one attaches a similar degree of credence to, there is a distinction to be made between a *w*-strong belief and the belief that *were P true* (or, false), then the agent would or will know it. This *subjunctive* epistemic state captures the confidence an agent has in her own beliefs. We model it via:

*Confidence*: '*A* knows that if *P* were (not) true, *A* would (not) know *P*': $Ak(P\epsilon T \rightarrow AkP)$. Confidence is a binary measure (it can be turned into a continuous measure if necessary) that can be used to answer the question: *Does A believe what she knows*? The converse question – i.e. *Does A know what she believes?* – is captured by *A's* awareness.

Alan sends an email message *p* to Betty. He also sends the message to Betty's boss, Charles, as a 'blind carbon copy' ('bcc'). In doing so, he creates the following epinet:

| | |
|---|---|
| Alan knows *p* | (Ak*p*) |
| Alan believes Betty knows *p* | (A*b*Bk*p*) |
| Betty knows Alan knows *p* | (BkAk*p*) |
| Alan believes Charles knows p | (AbCk*p*) |
| Charles knows Alan knows *p* | (CkAk*p*) – but, |
| Betty does not know Charles knows *p*; | (B~kCk*p*) |
| Charles knows Betty knows *p* | (CkBk*p*) |
| Charles knows Betty does not know Charles knows *p*; | (CkB~kCk*p*) |
| Charles knows Alan knows Betty does not know Charles knows *p* | (CkAkB~kCk*p*) – and |
| Betty does not know she does not know Charles knows *p* | (B~k(B~k(CkP))). |

Besides the asymmetry of information and potential mistrust of Betty he has seeded with Charles, the episode also gives Charles power over Alan that could be wield credibly by threatening to respond to Alan's message and sending a 'cc' copy of the response to Betty – on which the original 'bcc-to-Charles' will appear as a sub-header in the address field. Additionally, it matters a great deal whether Betty is *ignorant* of Alan's message to Charles (she does not know of the bcc, but knows it *may have happened* without knowing whether or not *it did happen*) – in which case she will look for additional signs and signals from both Charles or Alan – or, rather, if she is *oblivious* to the bcc, in which case she will not be vigilant.

### Epistemic network knowledge regimes

We are interested in laying out the epinets that common user behaviors on popular social platforms (e.g., Facebook, Twitter, Slack, LinkedIn) are most likely to induce, both among direct links ('friends', 'followers') and indirect links to whom some knowledge of direct ties is impacted – as is the case when Facebook profiles are not private. To do so, we need to introduce several higher-level distinctions among different knowledge regimes in epistemic networks, which relate to both direct (who knows what?) and interactive (who knows what is known by whom?) epistemic states, as follows [see Figure 2]:

*Distribution or Shared-ness*: 'A knows P and B knows P and C knows P and …': AkP & BkP & CkP & DkP &…. Distribution measures the spread of knowledge of *P* in a network *G* and can be measured in absolute (the total number of agents in *G* that know *P*) or relative terms (the proportion of agents in *G* that know *P*).

*Near-Commonality of Level n* ($NC^n$): 'A knows P, B knows P, A knows B knows P, B knows A knows P, and so forth, to level n': (AkP) & (BkP) & (AkBkP) & (BkAkP) &…, abbreviated: $(AkBk)^n(P)$ & $(BkAk)^n(P)$. Commonality of level *n* measures the level of interactive knowledge of *P* of actors in *G* and is a measure of the *coordinative potential* of the network. *Level-2* commonality (mutual knowledge), and level-3 commonality (AkBkAk(P) & BkAkBk(P)) of knowledge about some proposition *P* are particularly important for studying network-level mobilization and coordination.

*Covertness* is a property of shared, distributed or shared information in a subnetwork *SG* of a network *G*. Information in *SG* that is *covert* with respect to *G* is shared, mutual or common knowledge among members of *SG*, but is in the zone of *oblivion* of members of *G* that are not in *SG*. Examples range from military communications that are designed to look like noise to the receivers and potential eavesdroppers, as well as information communication networks in large corporate hierarchies in which top managers reach out 'covertly' to those who report to their reports, or their reports' reports, in order to discern if their direct reports are telling 'the truth'.

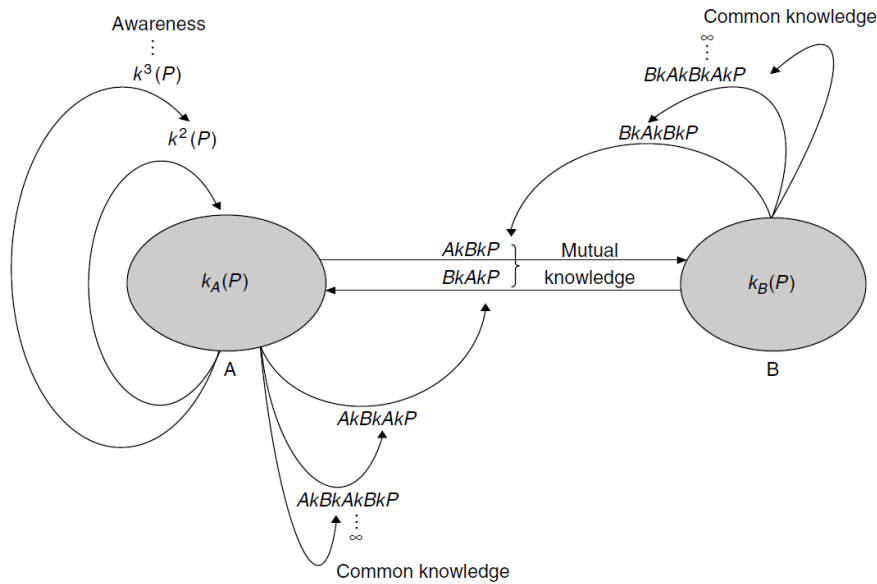

*Figure 2. Knowledge Regimes in a Dyadic Epinet.*

### Epistemic network knowledge neighborhoods

To achieve a characterization of epistemic structures in a network that serves our analysis of social platforms, we also require a representation of the i*nteractive knowledge neighborhoods* that result from the exchange of information over social platforms*:* groups of users that have knowledge (or, beliefs) about what other users know or believe. We can differentiate among [see Figure 3]:

*Distributed Knowledge Neighborhoods (N$_{DK}$(G))*: Subnetworks *SG* of network *G* whose members share knowledge of some proposition *P*. Every member of the network knows *P*, but, in a purely distributed knowledge neighborhood, no one knows that anyone (else) knows it.

*Mutual Knowledge Neighborhoods (N$_{Kn}$(G))*: Subnetworks *SG* of network *G* whose members share level-2 knowledge of P. Mutual knowledge neighborhoods describe networks in which *there is knowledge about the nature and extent of distributed knowledge*. Mutual knowledge undergirds subnetwork mobilization in situations characterized by 'I'll go if you go' scenarios (Chwe, 1999, 2000). Suppose *A* knows 'I'll go iff *B* goes' and *B* knows 'I'll go iff *A* goes'. *A* will not mobilize unless she also knows of *B* that 'he'll go iff she goes' and *B* will not mobilize unless he knows of *A* that 'she'll go iff he goes'. If *A* does indeed know that *B* will mobilize if and only if she mobilizes and B knows that *A* will mobilize if and only if he mobilizes, then mobilization can happen.

*Almost-Common-Knowledge Neighborhoods ($N_{SE}(G)$)*. Subnetworks *SG* of network *G* whose members have common knowledge of *level n* of P. 'Full common knowledge' neighborhoods can be used to model networks in which relevant knowledge *P* is self-evident, as in the case of a disclosure that occurs in a video call, in which everyone is co-present.

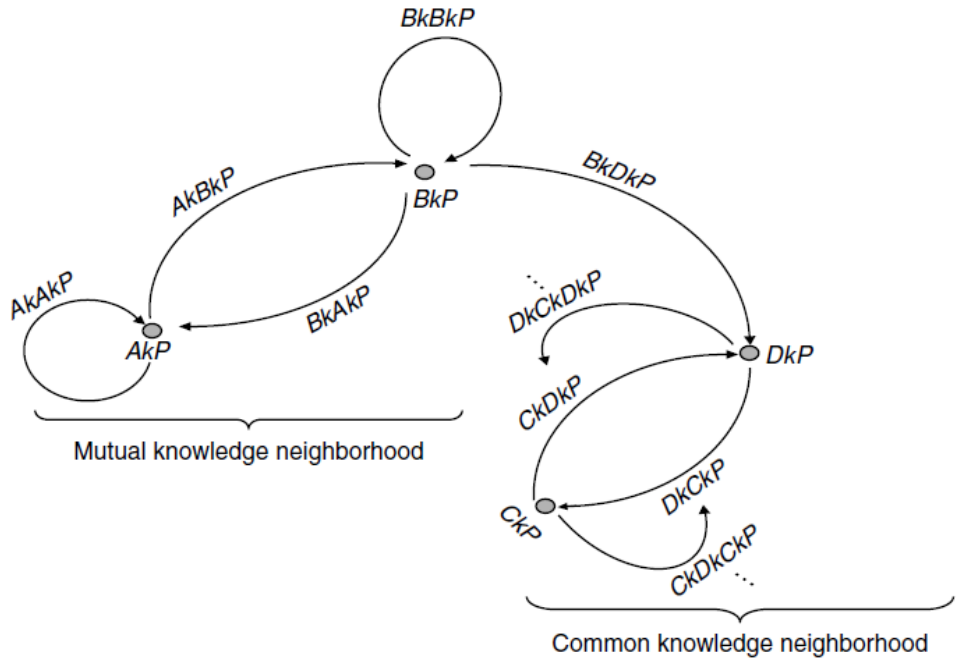

*Figure 3. Interactive Knowledge Neighborhoods in a Three Agent Epinet.*

### The Epinets that Grow on Social Platforms

Now we can apply epinet representations to the social structures engendered by massively-distributed social platforms. The epinets that characterize *social platforms* will depend on the specific decision rights users are given over basic actions (e.g., 'friend/unfriend', 'share video', 'like/dislike', 'comment', 're-tweet', etc.) and specific ways in which information about users is displayed and shared with other users (e.g., 'who sees I know whom?', 'who sees what I have shared with whom?' etc.).

### Epistemic state space of social platform-enabled actions

The epistemic state space of a social interaction platform depends on its basic design features. Consider:

<u>Email</u> and other point-to-point platforms easily generate distributed knowledge, but require additional steps to generate higher level knowledge and an infinite number of steps to generate fully common knowledge: In a dyad, before communicating, if *A* knows *P,* B will now know it until *A* emails *B* with a message conveying *P*. Once *B* receives *A*'s communication of *P*, *B* will know *P* and know *A* knows *P*. But unless *A* replies *B* will not know that *A* knows *P*. And so forth. Every reply produces a higher level of common knowledge, but fully common knowledge is only achieved in the limit as the number of communications tends to infinity.

WhatsApp/WeChat and other fast-messaging platforms emit a signal showing *A* that *B* has received her message [although this feature can be disabled]. Upon receipt of this signal, *A* will know that *B* knows *P,* even if *B* has not answered. And, knowing *A* will have received a notification of her having read *P*, *B* will know *A* knows *B* knows *P* as soon as *B* will have read *P*. Although common knowledge still requires the exchange of an infinite number of exchanges, mutual knowledge is quickly established via the acknowledgement signal. Importantly, whereas on point-to-point platforms the establishment of mutual knowledge is left to the recipient (who must choose to reply in order for the sender to know the recipient has received her message), the establishment of mutual knowledge in automatic-acknowledgement platforms is automated.

Twitter and other point-to-multipoint platforms enable rapid sharing of short (~200 character) messages that, upon 'tweeting', become public and distributed knowledge: anyone, with or without an account, can see them. But, unlike other platforms, the sender has no knowledge of who has seen the message. To be known as having seen or read a message, another user must like/dislike, comment or 're-tweet' the message to others.

### How social platforms build knowledge neighborhoods

There is much that social platforms *do* besides building knowledge neighborhoods, but *the difference that makes a difference* in the attractiveness of social platforms rests, to a large extent, on their ability to create knowledge neighborhoods of the *right kind* – i.e., which fulfill the social and relational needs of their members. For instance:

YouTube is a video-sharing and rating platform that allows individual members to upload and share videos they create with a broad and nameless community (*distributed knowledge neighborhood*: all users) and garner feedback from platform members who comment, 'like' or 'dislike' the videos (*mutual knowledge neighborhood*: commenters, who implicitly signal they know the primary user knows the shared information) and answer select comments to some commenters and (anonymous) raters (*mutual knowledge neighborhood*: subset of responders to whom the primary user responds).

Change.org is a social platform designed to engineer mutual knowledge neighborhoods that solve precisely the mobilization problem ('I will go if you go'). Members can initiate petitions for specific changes (ranging from the removal from office of a public official, magistrate, non-performing university president, dean or other bureaucrat, to the introduction of new local, municipal or provincial/state by-laws, to requiring or mandating local, municipal or state/provincial governments to resolve specific infrastructure problems). Circles of users who co-mobilize around a petition are 'movements' that can be initiated by a user (who argues for a particular measure) and is backed by 'supporters' or 'signatories' who, by signing, signal their support for the motion, disclose their identities as supporters, and come to be known as knowing and supporting the motion.

Twitter is a message broadcasting platform that allows the primary users ('Tweeters') to share thoughts and links with a large user base (which ranges beyond the base of subscribers of Twitter and generates a *distributed knowledge network*) and garner feedback in the form of 'likes', 'dislikes', or 're-tweets' – which create *mutual knowledge networks* – and subsequently expand the *mutual knowledge network* in the form of additional 'likes', 'dislikes' and 're-tweets'.

The creation of higher-level knowledge neighborhoods is constrained by the platform's limited interaction capabilities, which also limits the level of common knowledge that can be established among users.

Zoom and other video-conferencing platforms allow users to convene for real time or pre-scheduled video meetings (creating *common knowledge neighborhoods* among disclosures made in the video meeting). Additionally, agents who are co-present in a video meeting can publicly or privately message one another (creating *distributed, mutual and almost common knowledge subnetworks* of the people on the meeting); and, with the permission of participants, can record the meeting (who thereby come to know that others, potentially not at the meeting, will watch the meeting recording, thus creating a large *distributed knowledge neighborhood* the size and membership of which is unknown to the meeting participants). Or, *without* the explicit permission of anyone else attending the meeting, participants can use privately deployed screen recording software to store the contents of the meeting, which they can subsequently share with others without the permission of other meeting participants (which then creates a *distributed knowledge neighborhood* whose existence meeting participants will be *oblivious* of).

Quora is a 'question and answer' platform that builds user groups around questions posed by members and answers given by (presumably) other members in one of *M* 'interest areas' in which members (stipulate they) are interested. 'Topics' are *common knowledge* across the members of the platform. 'Questions' are *distributed knowledge* across the platform and *mutual knowledge* among members interested in the same topic. 'Answers' are *mutual knowledge* among members who belong to the same interest circles (and obviously among question posers and answer-writers). Breadth and depth of the generated epinets are established through additional moves Quora enables its members to make: Larger *mutual knowledge neighborhoods* ('I know you know this and you know I know this') are created by members who rate answers; and higher levels of *interactive knowledge* ('I know you know I know this...') can be established through comments on answers by other members – including the original question-askers.

Facebook is a 'personal information sharing and commenting platform' that allows users to share personal information (photos, videos, files, links, etc.) either with other, select users of the same platform who are 'trusted' (i.e., 'friends', creating *mutual knowledge neighborhoods*), or publically with all users (creating *distributed knowledge neighborhoods*). Facebook also allows detailed messaging among users that belong to the same circle (creating *almost-common knowledge neighborhoods*) or, enhanced communication to select users of the same circle who are singled out by a given user (creating higher *level almost common knowledge neighborhoods* that are not discoverable by other users, and even among users in the same circle).

### Trust as an Interactive Epistemic State

*Friends 'trust' friends* – as in the above Facebook example – and that trust undergirds the ability for members to share information freely. But what does trust mean in an epinet? Our definition makes use of modal epistemic structures such as the one we used to define *confidence*. As such, trust can be understood as a form of confidence. We decompose trust into two separate entities, *trust in integrity* (agents' propensity to say no more and no less than that which is true) and *trust in competence* (agents' propensity to know no more and no less than that which is true) – as follows:

*Trust in Integrity (Strong Form) ($T_I$)*: 'A knows that iff B knew P, B would assert P': $Ak(BkP \leftrightarrow BaP)$, where '*BaP*' denotes '*B* acts on *P*' or, '*B* asserts *P*', and *P* denotes a proposition that is relevant to the interaction between *A* and *B*. If *A* trusts *B*'s integrity, then *A* knows that 'iff *B* were to know the truth, then *B* would assert the truth'. In this form (a) knowledge by *B* of *P* is sufficient to guarantee to *A* that *B* would say *P* and (b) if *P* were false, then *B* would not say it, i.e., *B* speaks the truth and nothing but the truth.

This definition of trust generates the following interactive belief hierarchy between *A* and *B*: Suppose *P* is the proposition '*B* does not trust *A*'. Assume *P* is known to *B*, unknown to *A* and (self-evidently to both *A* and *B*) relevant to their interaction. If it is true that *A* trusts *B*, then *A* knows that iff *B* knew that *P* is true then *B* would assert it, and therefore that *B* would inform *A* of his mistrust. Because of the biconditional (iff) nature of the relationship between knowledge and expression, *A* can infer from the fact that *B* does not say 'I don't trust you' the fact that *B* trusts *A*. The argument is symmetric for *A*.

The second part of our definition of trust refers to *trust in competence*. *A* may trust *B*'s intentions, but not necessarily *B*'s ability to deliver, perform or make good on those intentions. In the way we have defined trust, *B* is only expected by *A* to say what he knows to be true, but not to know that which is true. We extend our definition of trust as follows:

*Trust in Competence ($T_C$)*: 'A knows that iff P were true, B would know P': $Ak(P\varepsilon T \rightarrow BkP)$. If *A* trusts in the competence of *B*, then *A* will trust that *B* is a faithful register of true propositions and not a register of false propositions, i.e., that *B* will register the whole truth.

Now we are in a position to characterize trust as the combination of trust in integrity and trust in competence. In the strong form, trust combines the strong form of trust in integrity (which we shall use henceforth unless we signal otherwise) with trust in competence:

*Trust Tout Court (T)*: A trusts in both the competence and integrity of *B*: $ATB \rightarrow (AT_I B \& AT_C B)$. If *ATB*, then *A* knows that *B* will assert the truth, the whole truth and nothing but the truth.

Moreover, for any or all of a collection of actors *A*, *B*, and *C*, the relation *T* has the following properties:

*Binariness B*. $\forall A, B : ATB$ or $\sim ATB$. 'For any *A* and *B*, *A* either trusts or does not trust *B*'. Proof: Focus first on $T_I$. Suppose that $AT_I B$ and $\sim AT_I B$, i.e., $Ak(BkP \leftrightarrow BaP) \& Ak(BkP \leftrightarrow \sim BaP)$. But this contradicts the binariness property of the k relation (a). The same argument applies for $T_C$ and therefore for the conjunction $T=T_I \& T_C$.

*Transitivity T*. $\forall A, B, C : ATB \& BTC \rightarrow ATC$. 'For any *A*, *B*, *C*, if *A* trusts *B* and *B* trusts *C*, then, *A* trusts *C*. Proof: If $AT_I B$, then $Ak(BkP \leftrightarrow BaP)$. If $BT_I C$ then $Bk(CkP \leftrightarrow CaP)$. Let $P='CkR \leftrightarrow CaR'$ for some *R*. Clearly, *BkP* and, *ex hypothesis*, *BaP*. Because *A* knows that *B* will only say what *B* knows (to be true) and knows to be true only those sentences that are in fact true, *AkP* and therefore $AT_I C$. Now, let $Q='R\varepsilon T \leftrightarrow CkR'$. Clearly, *BkQ* and, given $AT_I B$, $Ak(Q\varepsilon T)$. Therefore, $AT_C C$. Therefore $AT_C C$ and $AT_I C$, i.e., *ATC*.

Does the fact that *A* trusts *B* entail the fact that *B* trusts *A*? The answer is 'no'. The proof is by counter-example. Suppose *A* trusts *B* to assert 'the truth, the whole truth and nothing but the truth', but *B* does not trust *A* to assert the truth, the whole truth and nothing but the truth. Let *P* represent the proposition '*B* does not trust *A* to assert the truth, the whole truth and nothing but the truth'. *Ex hypothesis*, *P* is true. Since *A* trusts *B*, *B* will assert *P* and *A* will know *P*, and therefore will know that *B* does not trust her. So, if *A* trusts *B* and *B* does not trust *A* then our definition requires *A* to know that *B* does not trust *A*. However, in order for *A* to not trust *B* as a result of this knowledge, it is necessary to make *A*'s trust in *B* conditional upon *A*'s knowledge of B's trust in *A*, which it does not necessarily have to be: *A* may trust *B* on account of the 'kind of person *A* believes *B* is' rather than on account of a set of expectations about *B*'s behavior given *B*'s incentives.

### Epistemic network trust regimes

This characterization of trust allows us to identify sufficient conditions for the knowledge that trusting, trusted, secure, and authenticated networked actors fulfill. In particular, we define the following:

*Trust neighborhood $N_{Ti}(G)$, $N_{Tb}(G)$*: Fully-linked subnetwork of *G* (clique) that shares trust in mutual competence and integrity (weak form). A trust neighborhood is a good model for a network of 'close ties', wherein actors can rely on one another for truthful and truth-like knowledge sharing, conditional upon awareness: if an actor is aware of *P* (knows it and knows she knows it), then she will share *P* with others. Thus, in a trust neighborhood, communication is truthful, but what is being communicated may not be 'the whole truth'. Trust neighborhoods can be used to represent referral cliques, often used to get the 'inside story' on people and organizations that have a history of interactions within an industry. A referral clique is a clique of actors in which information flows are 'trustworthy' in the technical senses of trust that we introduced earlier. Within a trust neighborhood (a 'circle of trust') sensitive information is likely to flow more reliably and accurately than outside of it. The epistemic approach we have taken to representing trust allows us to map the precise trust neighborhoods within a network, and therefore to make predictions about the reliable spread of accurate information within the broader network of contacts.

*Security neighborhood $N_S(G)$*: Fully-linked subnetwork of *G* (clique) that shares trust in mutual competence and integrity (strong form). Security neighborhoods are high-trust cliques – and may be good representations for networks of field operatives in law enforcement scenarios, where authentication of communicated information is crucially important to the payoff to each actor, or conspiratorial networks, for example, a subgroup of a board of directors trying to oust in the company's CEO. The reason for calling such a network a security neighborhood is that it has (common) knowledge of itself *as* a trust neighborhood, and thus possesses an important authentication mechanism for communication, which trust neighborhoods based on weak-form trust do not possess. If *A* and *B* belong to the same trust neighborhood, then *C* (a third party unknown to *A* but known to *B*) can interject herself in the communication between *A* and *B* and contribute information to *A* and *B*. If the fact that C is part of the trust neighborhood is not itself common knowledge, then the information that *C* contributes is not authenticated, in the sense that it is not possible for *A* to decide without consulting *B* whether or not to trust information coming from *C*. If, on the other hand, *A* and *B* are part of a security neighborhood, then *C* will immediately be recognized as an outsider. Common knowledge of clique membership is a key element in the process of authentication, which is itself important in subnetworks concerned with infiltration from the outside. A key property of a security neighborhood is thus that it necessarily has common knowledge of the fact that it is a trust neighborhood:

Proposition: A security neighborhood is a common knowledge neighborhood.

Proof: Consider a two-person clique (*A*, *B*) where *ATB* and *BTA*, both in the strong sense. For any *P*, it is the case that *Ak('P='true'↔BaP)* and *Bk(P='true'↔AaP)*. For instance, let *P='ATB'*. Since *P* is true and *Ak('P='true'↔BaP)*, *Ba(ATB)* is true. Now, suppose *P* is true, but there is some level of almost-common knowledge, *n*, at which it is not the case that $(AkBk)^n P$, i.e., $\sim(AkBk)^n P$. Then, at almost-common knowledge level, *n-1*, it cannot be the case that $(AkBk)^{n-1}P$ (to see this, let $P^{n-1}$ represent the proposition '$(AkBk)^{n-1}P$', which is true, and together with *ATB* implies $(AkBk)^n P$). Therefore, if *P* is true, *ATB* and *BTA*, then *P* is common knowledge.

Trust matters crucially to the way in which any *ego* conveys information to an *alter*. Therefore, classifying social network ties in terms of the degree of trust that linked actors share should allow us to make progress in understanding the dynamics of critical information in a network. Using trustful-trusting relationships as building blocks, we define the most likely, fastest, or most reliable paths by which relevant and truthful information will flow in a network:

<u>Trust Conduit</u> from actor *A* to actor *J*: path *A-B-C-…-J* from *A* to *J* passing through actors *B,…,I* such that *A* trusts *B*, *B* trusts *C*, *C* trusts *D*, …, *I* trusts *J*. Trust conduits can enable reliable knowledge flows in networks: information that comes from *A* will be trusted by *B*, information that comes from *B* will be trusted by *C*, and so forth, such that information flows credibly along a trust conduit. Trust conduits can represent knowledge pipes in organizations and markets, and thus to study the dynamics of the propagation of new information. They can also be used to distinguish between facts and rumors: facts are bits of information that have propagated along a trust conduit, whereas rumors are bits of information that have not. Since facts can be used to check rumors, trust conduits not only enable speedy propagation of useful relevant information, but also to provide checks and constraints on propagation of rumors. Rumors should thus die out more rapidly in networks seeded with many trust conduits than those lacking them.

<u>Trust Corridor.</u> Because trust is not in general symmetric (particularly in its weak forms), we define a trust corridor as a two-way trust conduit. A trust corridor is useful for representing reliable bi-directional knowledge flows within a network, thus increasing the degrees of freedom associated with any particular flow. If reliable knowledge can flow in both directions in a knowledge pipeline, rumor verification can proceed more efficiently, as any one of the actors along the path of the corridor can use both upstream and downstream actors for verification purposes. Trust corridors both accelerate the reliable transmission of facts and also to impede the promulgation of unverified or unverifiable rumors within a network.

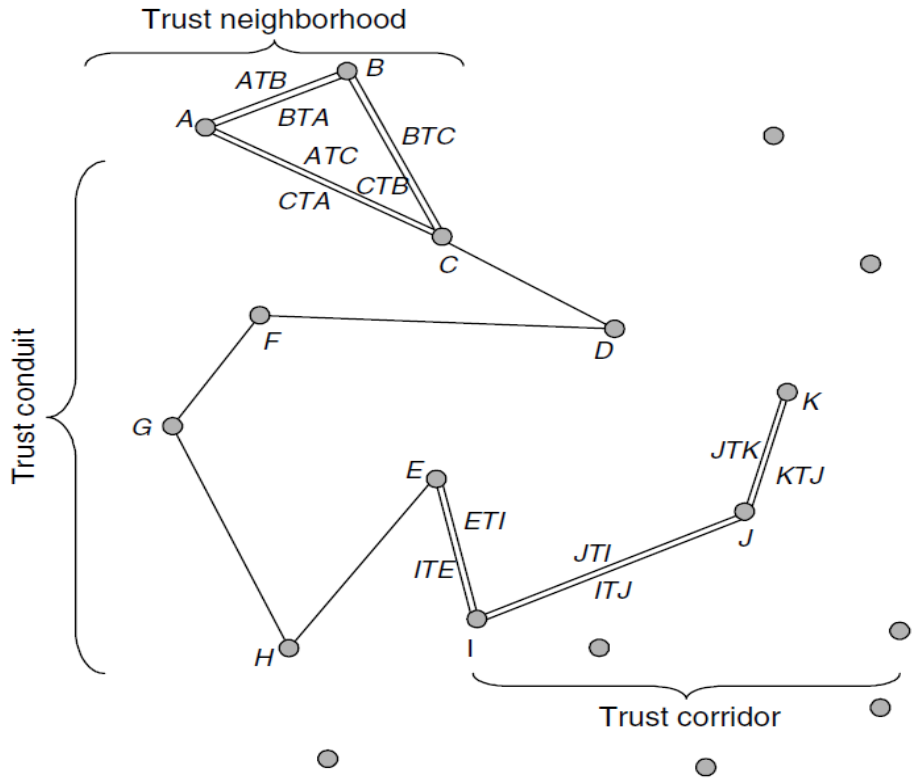

*Figure 4. Trust neighborhoods and corridors in an epinet.*

Special relationships are required for the propagation of "sensitive" information – information that senders wish to be assured reaches all and only intended recipients. In such situations, actors' knowledge of the integrity of the trust conduit they access is required for the conduit to function as an information channel. The conditions for the existence of such conduits can be made more precise by the use of epinets, as follows:

Security Conduit from actor *A* to actor *J* is a path *A-B-C-...-J* from *A* to *J* passing through actors *B,…,I* such that *A* trusts *B*, *B* trusts *C*, *C* trusts *D*,…, *I* trusts *J* and the conduit is common knowledge among *(A,…,J)*. This feature renders them robust against infiltration, as each user of the communication channel embodied in the conduit can be authenticated by any of the actors comprising the conduit. Security conduits are authenticated knowledge pipes in which rapid authentication of incoming messages can be performed quickly using actors' (common) knowledge of each other's membership in the conduit.

Security Corridor is a two-way security conduit. It can be used to represent subnetworks in which (a) rumor verification can be performed more rapidly than in one-way knowledge conducting structures, and (b) the probability of undesired 'infiltration' is low because of the authentication mechanism proper to 'strong form' trust-based subnetworks. Security corridors may also be used to represent authenticated expert networks which are characteristic, for instance, of robust social referral networks which can be used both for fast rumor verification and the robust circulation of reliable knowledge.

### How social platforms build trust and security neighborhoods

As we saw in the case of first- and higher-level knowledge neighborhoods, social platforms can seed and help crystallize epinets by virtue of their information conveyance, sharing and diffusion properties, and of the decision rights they give users for linking up with other users and sharing and un-sharing information with them. To wit:

LinkedIn is a professional networking platform that allows users to store and share their credentials, skills, endorsements and written contributions, and to connect with other users, endorse them, search for users in particular fields and professions, and attract other users to connect with them via the information they display. The fact that (premium) members can see all of the members that have looked up their profile automates the creation of mutual knowledge neighborhoods ('I see you saw my profile and know you know the information it contains') and, if the member who looks up the information knows the primary user is a premium user, drives mutual knowledge towards higher levels of knowledge ('You know I know you know the information in my profile.') Although it does not measure trust in the explicit way in which we have defined it, the platform has a link architecture that requires some level of trust for accepting an invitation, as all of the acceptor's and inviter's contacts are instantly shared among them once an invitation is accepted. A user's 'trust neighborhood', thus, becomes all of the direct contacts of the user. 'Premium' versions of the platform allow users to see the connections of the connections of their connections, thus broadening the trust neighborhood.

Slack is a multipoint-to-multipoint messaging and information conveyance platform that allows users to form dedicated channels of communication that include certain subsets of users engaging in regular, private or covert communication on a regular basis. Dedicated private channels obviate the need for deciding on a recipient list (and cc and bcc lists in the case of email) every time a member wishes to communicate. Membership in such channels also involves a form of trust – in particular, the trust that all communication that is relevant to a particular topic will occur within the confines of the channel (i.e., that a member will not unilaterally email or reach out to other members of the channel on a topic of interest to all channel members, or that someone will accidentally or intentionally be 'dropped' from a communication to other members of the channel. (Or, in the counterfactual analysis of trust we have laid out: that if it were the case that an *outside-of-channel* communication were to occur, that communication would be reported to members of the channel). *Covert* channels are not visible to other members of the same Slack organizational account (and as such become security neighborhoods or conduits, in which the common knowledge members have of the trust and secrecy covenants safeguard the covertness of the communications).

### Discussion: Epinets as an Analysis and Design Tool

The value of a new representation tool is tested by the new applications or modifications it enables – as much as it is by the understanding of user behaviors it confers. To put epistemic networks to a pragmatic tests, we ask: what changes and modifications to existing platforms would an epistemic analysis and understanding of *any* social platform *qua* network-generating entity enable or suggest?

<u>Automating mutual knowledge discovery</u>. For broadly distributed-knowledge platforms (e.g., YouTube or Twitter), automating the creation of mutual knowledge neighborhoods (entailing that authors could immediately see everyone who has seen their video or read their tweet, as opposed to simply the number of people that have viewed it) can increase the level and specificity of user interaction.

<u>Increasing network visibility for premium users</u>. For platforms that automatically create common or mutual knowledge neighborhoods (e.g., Facebook, Instagram or LinkedIn), users' 'network visibility' is limited to other members that are at most two (friends' friends: Facebook) or three (contacts' contacts' contacts: LinkedIn Premium) links away. However, given the commutative nature of trust, it may be the case that if *A* trusts *B*, who trusts *C* who trusts *D* who trusts *E*, then *A would* trust *E*, provided that a 'friendship' or 'contact' are also trust relationships.

This may require creating a special class of ties or connections ('trusted'), which a platform can do by:

<u>Making trust covenants explicit</u>. The definition of trust we formulated can be used to make it clear to members of a platform what a privileged tie, like trust, entails, and therefore to differentiate between different kinds of ties. On the anticipation that very large trust neighborhoods can invite trust-breaking, free-rider behavior, one can also:

<u>Make trust breaches discoverable</u>. For example, by having the communication of information that is mutual or common knowledge in a trust neighborhood by a user to someone outside that neighborhood become distributed within the neighborhood itself.

<u>Create longer conduits of knowledge sharing, trust and security</u>. Social platform providers have access to the *entire* social networks their platforms spawn: they can see *all* members and all links among them, as well as all of the content that members upload, store, distribute and share with other members. They can *use* this knowledge, for instance (e.g., LinkedIn, Facebook) to inform members of the *fastest routes*, or social network geodesics [LinkedIn gives some path length information (1st, 2nd, 3rd, 3rd+) but not routes.], they can use to reach a member they are seeking a connection to, using trust conduits or corridors, for instance (which, as per our suggestion above to make trust ties explicit, are connections or ties that are privileged relative to a regular connection ('I trust X', for instance, can designate such special links). Several such links can be concatenated together (since $ATB\&BTC \rightarrow ATC$) to create trust conduits and corridors linking most members to most other members, which solves the 'social network navigation problem' that most members grapple with and many platforms try to solve, albeit inefficiently, with 'premium' subscription packages.